\documentclass[apj]{emulateapj}
\usepackage{epsfig}
\usepackage{epstopdf}
\usepackage{graphics}

\usepackage{xcolor}

\begin{document}

\title[]{The Galactic Center: A PeV Cosmic-Ray Acceleration Factory}
\author{Yi-Qing Guo$^1$, Zhen Tian$^1$, Zhen Wang$^1$, Hai-Jin Li$^{2,3}$, Tian-Lu Chen$^{2,3}$}
\affil{1 Key Laboratory of Particle Astrophysics, Institute of High Energy Physics, Chinese Academy of Sciences,Beijing 100049,China}
\affil{2 Physics Department of the Science School, Tibet University, Lhasa 850000, China}
\affil{3 The Key Laboratory of Cosmic Rays, Ministry of Education, Lhasa 850000, China}

\begin{abstract}
  The multi-TeV $\gamma$-rays from the Galactic Center (GC) have a cutoff 
  at tens of TeV, whereas the diffuse emission has no such cutoff, 
  which is regarded as an indication of PeV proton acceleration by the HESS experiment.
  It is important to understand the inconsistency and study the possibility that PeV cosmic-ray acceleration
  could account for the apparently contradictory point and diffuse $\gamma$-ray spectra.
   
  In this work, we propose that the cosmic rays are accelerated up to $>$PeV in GC. 
  The interaction between cosmic rays and molecular clouds is responsible 
  for the multi-TeV $\gamma$-ray emissions from both the point source and diffuse sources today.
  Enhanced by the small volume filling factor (VFF) of the clumpy structure, the absorption of the $\gamma$-rays 
  leads to a sharp cutoff spectrum at tens of TeV produced in the GC. 
  Away from galactic center, the VFF grows and the absorption enhancement becomes negligible.

  As a result, the spectra of $\gamma$-ray emissions for both point source and diffuse sources can be successfully reproduced
  under such self-consistent picture.
  In addition, a ``surviving-tail" at $\sim$100 TeV is expected from the point source,
  which can be observed by future projects CTA and LHAASO.
  Neutrinos are simultaneously produced during proton-proton (PP) collision. 
  With 5-10 years observations, the KM3Net experiment will be able to detect the PeV source according to our calculation.
\end{abstract}
\maketitle

\section{Introduction}
  It is well known that the GC, with a supermassive black hole ($\sim$4 $\times$ $10^6 M_\sun$),
  is a good laboratory for the study of astrophysical phenomena. Historically, there have been many
  discussions on the possibility that the GC is a dominant source of galactic cosmic rays 
  \citep{1981AZh....58..959P,1981ICRC....2..344S,1983JPhG....9.1139G,2013NJPh...15a3053G,2013JPhG...40f5201G}.
  With state-of-art technologies, current $\gamma$-ray observations provide unprecedented sensitivity in studying the 
  acceleration activities in the GC.

  Very high energy $\gamma$-rays from hundreds of GeV to tens of TeV in 
  the direction of the GC have been observed 
  by several atmospheric Cerenkov telescopes, such as CANGAROO 
  \citep{2004ApJ...606L.115T}, VERITAS \citep{2004ApJ...608L..97K,2015arXiv150806311S}, 
  HESS \citep{2004A&A...425L..13A,2006Natur.439..695A,2006PhRvL..97v1102A,2008A&A...492L..25A}, 
  and MAGIC \citep{2006ApJ...638L.101A}. 
  Later observations by HESS found the source spectrum has
  an exponential cutoff at about tens of TeV with the implication of 
  intrinsic origin \citep{2009A&A...503..817A}. This implies that the maximum accelerated energy
  for a proton is $\sim$200 TeV as shown \citep{2013JPhG...40f5201G}.
  The diffuse $\gamma$-ray emission is also observed at GC
  disk range by the HESS experiment \citep{2006Natur.439..695A}. The more interesting thing is that
  the $\gamma$-ray emission is correlated with the density of molecular hydrogen, which is
  generally regarded as a hadronic source. Simultaneously, the spectrum for the GC point source is
  the same as the diffuse one, and they may possibly share the same origin: the GC supermassive black hole.
  Just recently, the diffuse $\gamma$-ray emissions around the GC have been observed by the HESS 
  experiment \citep{2016Natur.531..476H}. The results support that the $\gamma$-ray emissions come from
  $\sim$PeV energy proton and the most plausible accelerator is the GC \citep{2016Natur.531..476H}.
  Several models have been proposed to explain the $\gamma$-ray emission and discuss the
  PeV acceleration at the GC region 
  \citep{2016arXiv160400003F,2016arXiv160408791C}.
  The direct criterion to PeV CR acceleration in the GC region is the observation of high-energy neutrinos.
  Several 30 TeV to 2 PeV neutrinos have been observed from the GC direction by the IceCube experiment 
  \citep{2013PhRvL.111b1103A,2013Sci...342E...1I}. Some works have been discussed the possibility of
  the GC origin \citep{2014PhRvD..90b3010A,2014PhRvD..90f3012B,2015ApJ...806..159K,2015PhRvD..92b3001F}.

  The problem is how to understand the cutoff in the spectrum of the central source. 
  One possible reason is the absorption of $\gamma$-rays
  by interactions with the ambient infrared radiation field. 
  But calculations showed that the absorption effect is not sufficient \citep{2009A&A...503..817A}. 
  It is possible that the 
  absorption of $\gamma$-rays is underestimated because 
  the infrared radiation field near the GC may be irregular.
  As a matter of fact, the material in the GC region is clumpy, dense, and fragmented \citep{2011AJ....142..134E}.
    The degree of irregularity can be described by the VFF, which is defined as
    the ratio of the volume of clumpy structure to the total volume of the GC. When the VFF is much smaller, 
    the material density of the clumpy structure is much higher than the fixed total material. 

  The VFF in circumnuclear disk is at the level of
  1$\%$ \citep{2001A&A...377.1016V,2002A&A...388..128V,2007ApJ...659..389F}.
  At the central cavity \citep{1993ApJ...402..173J,1987ApJ...318..124G},
  the gas density is large
  enough for self-gravity to form a clumpy structure to overcome the strong tidal shear of the black hole,
  and this will make the VFF even smaller, $\sim 0.1\%$
  \citep{1993ApJ...402..173J,1985ApJ...297..766G}.
  One important consequence is that the infrared
  radiation component of interstellar radiation field (ISRF) should have a VFF similar to the gas material.
  The reason is that the infrared background light comes from the reemitting of the gas after absorbing
  the starlight. A small VFF means that the $\gamma$-rays experience much 
  many more background photons
  being generated or passing through the dense gas region. That causes a
  much stronger absorption and attenuation at high energy. So the observed $\gamma$-ray 
  cutoff at tens of TeV can possibly be due to the attenuation of the ISRF.
  Away from the GC, the VFF grows, the absorption will become less and less important. 

  In this work, we propose that the CR could have been accelerated to $\sim$PeV during the 
  GC activity in past and are producing the high-energy $\gamma$-rays by $PP$-collision today.
  We further suggest that observed the sharp cutoff $\gamma$-ray spectrum is due to the absorption of the ISRF 
  enhanced by the dense clumpy structure in the GC. Considering the density of the ISRF
  and absorption efficiency,
  the higher-energy $\gamma$-ray around 100 TeV can escape and the surviving tail is predicted, 
  which can be tested by future projects, such as CTA and LHAASO experiments.
  Simultaneously, neutrinos
  can be produced during the $PP$-collision and can be observed 
  by the KM3Net experiment in a few years of operation.
  The paper is organized in the following way. In Sec. 2, we present the detailed modeling of this picture. 
  Sec. 3 is the discussions and Sec. 4 gives the conclusions.

\section{Model and results}

  During the violent activities,
  the accretion of stars and gas by the supermassive black hole could be
  effective in accelerating particles.
  The maximum energy that protons can achieve for diffusive
  shock acceleration is \citep{2005ApJ...619..306A}
\begin{equation}
    E_{\rm max} \sim eBR \approx 10^{14}\left(\frac{B}{\rm G}\right)
    \left( \frac{M}{4\times 10^6M_\odot}\right)\left(\frac{R}{\rm 10R_g}\right)\ {\rm eV}
 \end{equation}
  where $B$ is the magnetic field and $R$ is the size of the acceleration
  region. As in \citep{2005ApJ...619..306A}, we assume the acceleration
  takes place within $10$ Schwarzschild radii ($R_g\sim10^{12}$ cm) of the
  black hole. To accelerate protons to above $\sim$PeV requires magnetic field strength of
  tens of G in the acceleration region \citep{2009ApJ...698..676D,2001A&A...379L..13M}. 
  Such a condition could be
  reached in the very central region of the GC \citep{2005ApJ...619..306A,2013Natur.501..391E}.
  On the other hand, if the acceleration takes place in larger regions,
  the required magnetic field could be smaller. 
  When the accelerated CRs diffuse out of the GC, the hadronic interaction with interstellar medium (ISM) will occur
  and produce similar amounts of $\gamma$-rays and neutrinos. 
  The detailed model calculations are discussed
  in the following.

\subsection{The $\gamma$-ray emission in the GC with the break spectrum of protons in high energy}

\begin{figure*}[!htb]
\centering
\includegraphics[width=0.47\textwidth]{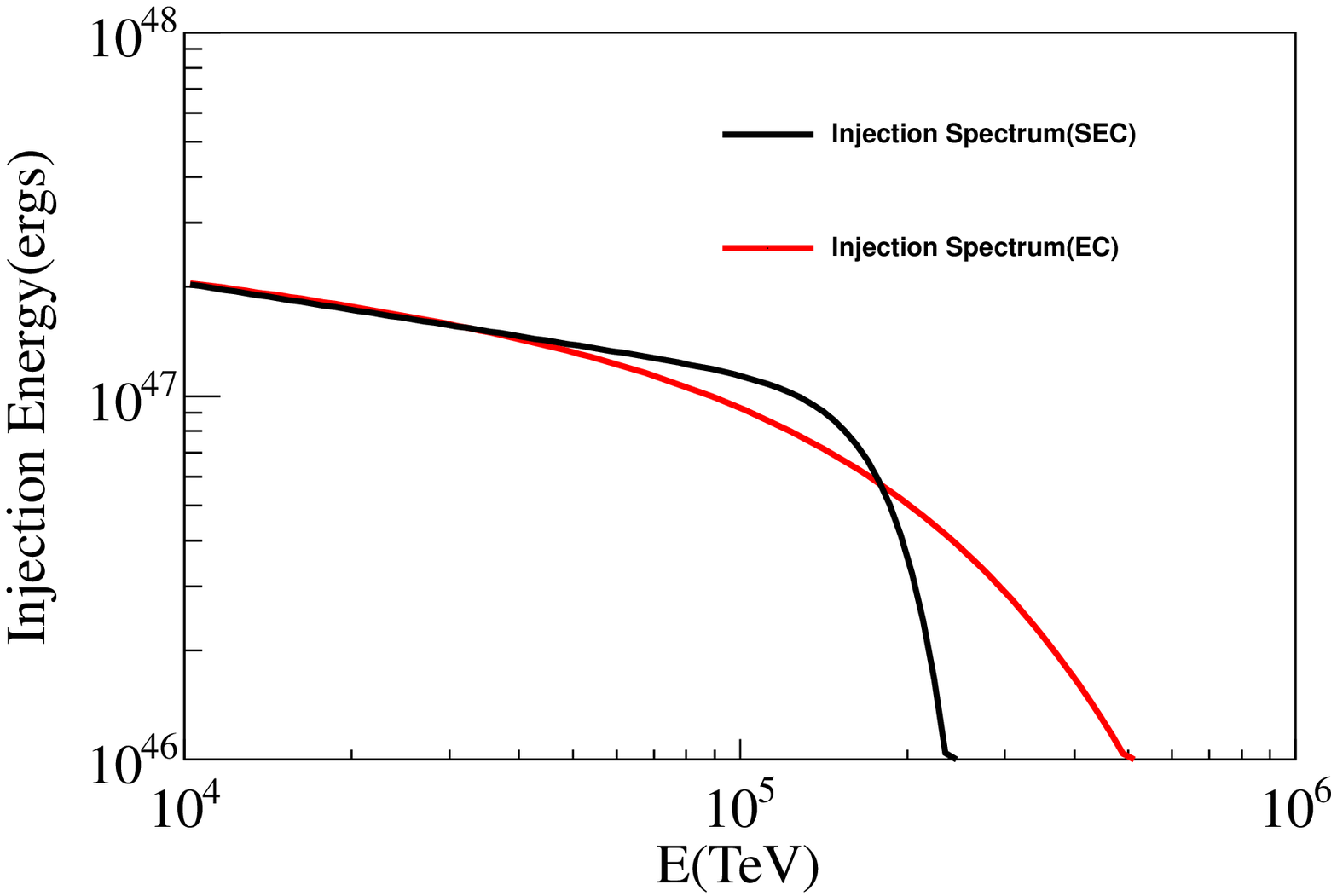}
\includegraphics[width=0.47\textwidth]{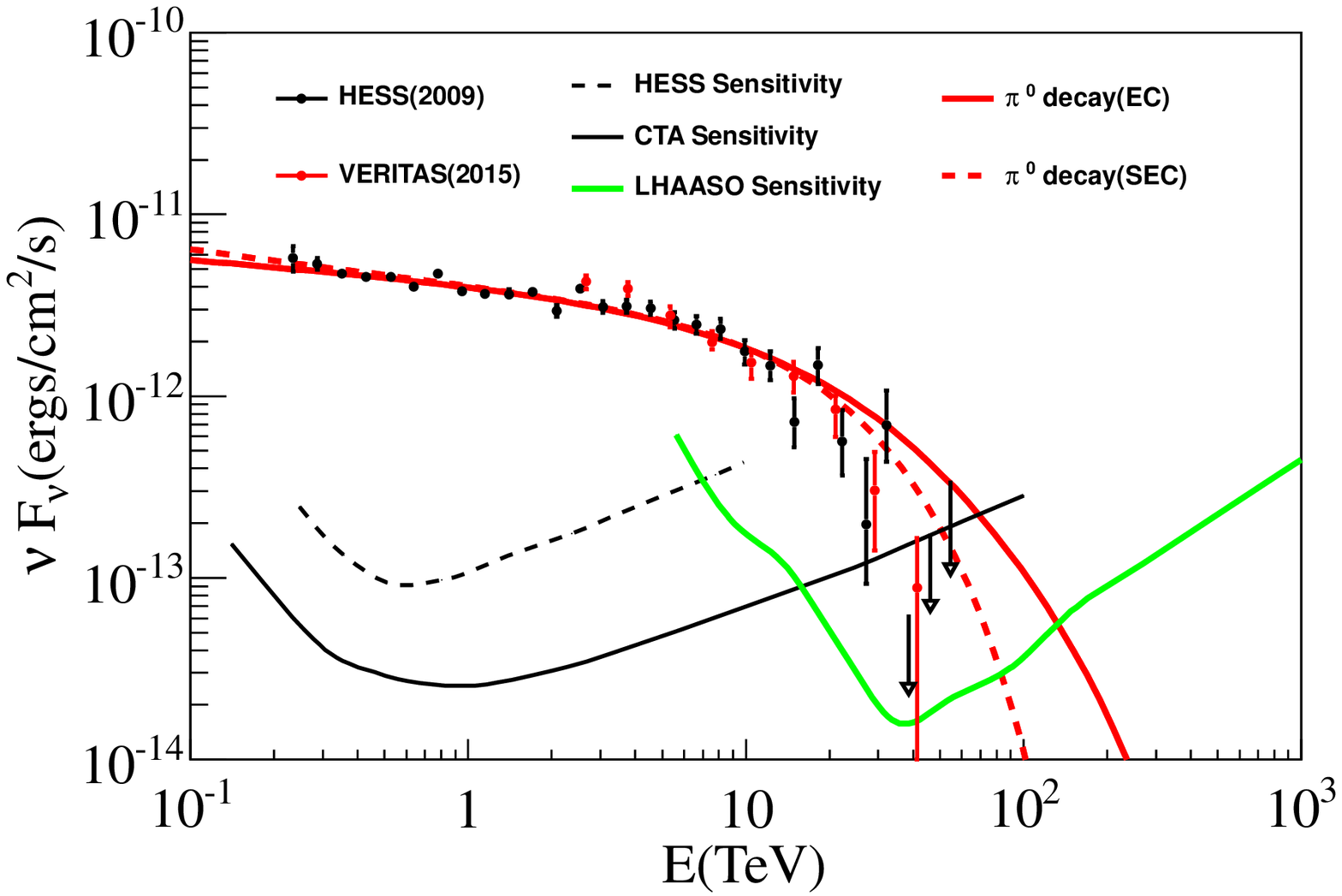}
\caption{Left: the injection spectrum of proton with an EC or SEC
         at critical energy $E_c=200$ TeV. Right:The $\gamma$-ray emission comparison between model calculation of $PP$-collisions and
         HESS observation \citep{2009A&A...503..817A}.}
\label{acellerateMax}
\end{figure*}

  The $\gamma$-ray emission from the point source in the GC
  has a broken power law spectrum at tens of TeV. The best fit of the cutoff can be described 
  by exponential function in high energy \citep{2009A&A...503..817A}.
  While adopting the traditional model of the ISRF, the absorption effect is too small to 
  explain the observed cutoff spectrum of HESS J1745-290 \citep{2009A&A...503..817A}. 
  The alternative solution 
  attributes it to the intrinsic cutoff, which characterizes the acceleration limit of 
  the flaring event. For comparison, the broken spectrum of protons can be simply adopted to exponential 
  cutoff (EC) as $e^{-E/E_c}$ or superexponential cutoff (SEC) as $e^{-(E/E_c)^s}$,
  where E is the proton energy, $E_c=200$ TeV is the critical energy and $s>>1$ denotes the sharp break.
  The key points are the
  density distribution of CRs and the ISM distribution in the GC region. For sake of simplicity,
  the average density $n_{\rm gas}$=$10^3 cm^{-3}$ is assumed in the GC point and diffuse 
  regions \citep{2011ApJ...726...60C,2012ApJ...753...41L}.
  The total energy of the CRs is dependent on
  its spectral index. Due to the energy break at 200 TeV, the spectrum of CRs becomes soft at tens of TeV. 
  Here the spectral index 2.15 (2.24) and total energy of 0.86$\times 10^{48}$(1.86$\times 10^{48}$) erg is 
  adopted for the EC (SEC) mode.
  Under such scenario, the spectrum of $\gamma$-rays are calculated as shown in Fig.\ref{acellerateMax}. 
  From this figure, it is clear that the proton spectrum with SEC is much better to fit the observation. 

  Although the $\gamma$-ray emission in the point source of the GC can be explained by adopting the 
  SEC of injection CRs, it is hard to reproduce the diffuse one around GC region under the same scenario. 
  The alternative method, like the absorption in the heavy ISRF, should be considered to understand the possible physical mechanism
  in one unified way. 

\subsection{$\gamma$-ray absorption with an inhomogeneous ISRF in the GC}

  The Galaxy is not transparent to very high energy $\gamma$-rays. 
  The main three processes resulting in energy losses of photons are the photoelectric
  effect, Compton scattering and pair production. The photoelectric effect and Compton scattering are
  negligible for the $\gamma$-ray with the energy higher than tens of TeV \citep{2014ApJ...795..100G}. So 
  the dominant contribution to the attenuation 
  comes from pair production, which leads to the change in the $\gamma$-ray spectrum.
  In this work, the absorption can be divided into two components: within the source region
  and on the way from the source region to the Earth. For the latter, 
  previous studies
  \citep{2006A&A...449..641Z,2006ApJ...640L.155M} have shown that
  the absorption is just 10$\%$ for 20 TeV $\gamma$-rays and 20$\%$ for 50 TeV $\gamma$-rays,
  which is far less than what is required in order to explain the cutoff spectrum of the point source at the GC \citep{2009A&A...503..817A}.
  Absorption in the source region might be more complicated and need special consideration.

  The energy-dependent absorption of $\gamma$-ray can be described as $e^{-\tau(E)}$, 
  where $\tau(E)$ is the optical depth for $\gamma$-ray in energy E. Similar to previous work \citep{2006A&A...449..641Z,2006ApJ...640L.155M},
  $\tau(E)$ can be described in the source region as
\begin{eqnarray}
\tau(E) & = & \int _{R_0} dr \int d\cos(\theta) \int 
    \frac{dn(\epsilon,r)}{d\epsilon} \nonumber \times\\
    & & \sigma_{\gamma\gamma}(E,\epsilon,\cos\theta)
    \frac{1-\cos\theta}{2} d \epsilon,
\label{eqEBL}
\end{eqnarray}
  where $\epsilon$ is the energy of the ISRF photon, 
  $\sigma_{\gamma\gamma}$ is the pair 
  production cross section and can be precisely derived \citep{1967PhRv..155.1404G}. Then the attenuation is only dependent on the
  the differential number density $dn(\epsilon,r)/d\epsilon$ and the size of the ISRF region $R_0$ with the value of 2 pc.

  The average photon intensity at the far-infrared band from the GC region 
  has been measured by Herschel PACS and SPIRE \citep{2011AJ....142..134E} and can be 
  defined as $I_\epsilon ~(photons~ s^{-1} cm^{-2} sr^{-1})$. 
  In the traditional way, the attenuation of $\gamma$-rays is calculated by adopting
  this homogeneous radiation field.
 In fact, the radiation field can be described by the point source formula,
  as $I_\epsilon \propto 1/r^2$.
  So the photon density in the GC region can be approximately estimated as 
  $dn(\epsilon,r)/d\epsilon = \frac{4\pi}{c}\cdot I_\epsilon \propto \frac{4\pi}{c} \cdot 1/r^2$. 
  However, the radiation field is very clumpy with a VFF (denoted as $f_V$). We can consider the effect of 
  the clumpy structure by replacing $r \rightarrow r\cdot f_V^{1/3}$. Under such situation,
  the photon density in the clumpy structure should be
  corrected to $dn(\epsilon,r)/d\epsilon \sim\frac{4\pi}{c}I_\epsilon/f_V^{2/3}$ and the corresponding 
  integration of $dr$ is the radius of the clumpy structure as $R_0\cdot f_V^{1/3}$.
  In this work, the attenuation of $\gamma$-rays is calculated by adopting this enhanced differential 
    number density $dn(\epsilon,r)/d\epsilon$.
  So compared with the traditional calculation, the absorption should be enlarged by a factor of $1/f_V^{1/3}$ after considering
  the VFF $f_V$ based on Equation \ref{eqEBL}. 
  One special case of our model is that the $\gamma$-rays are produced in the very center of the GC
  and the attenuation is similar to the work \citep{2005ApJ...619..306A,2016arXiv160408791C}.  

\begin{figure*}[!htb]
\centering
\includegraphics[width=0.47\textwidth]{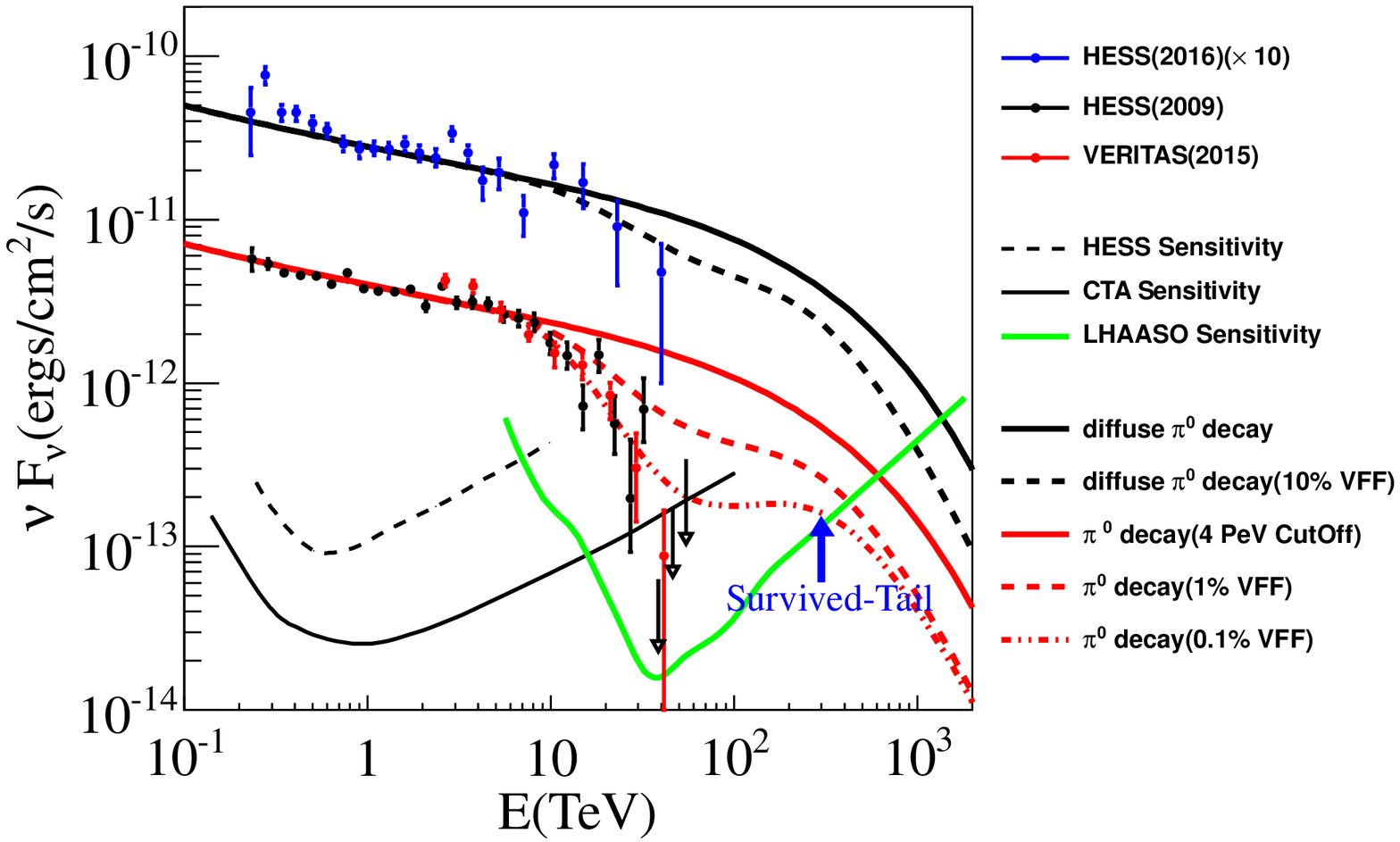}
\includegraphics[width=0.47\textwidth]{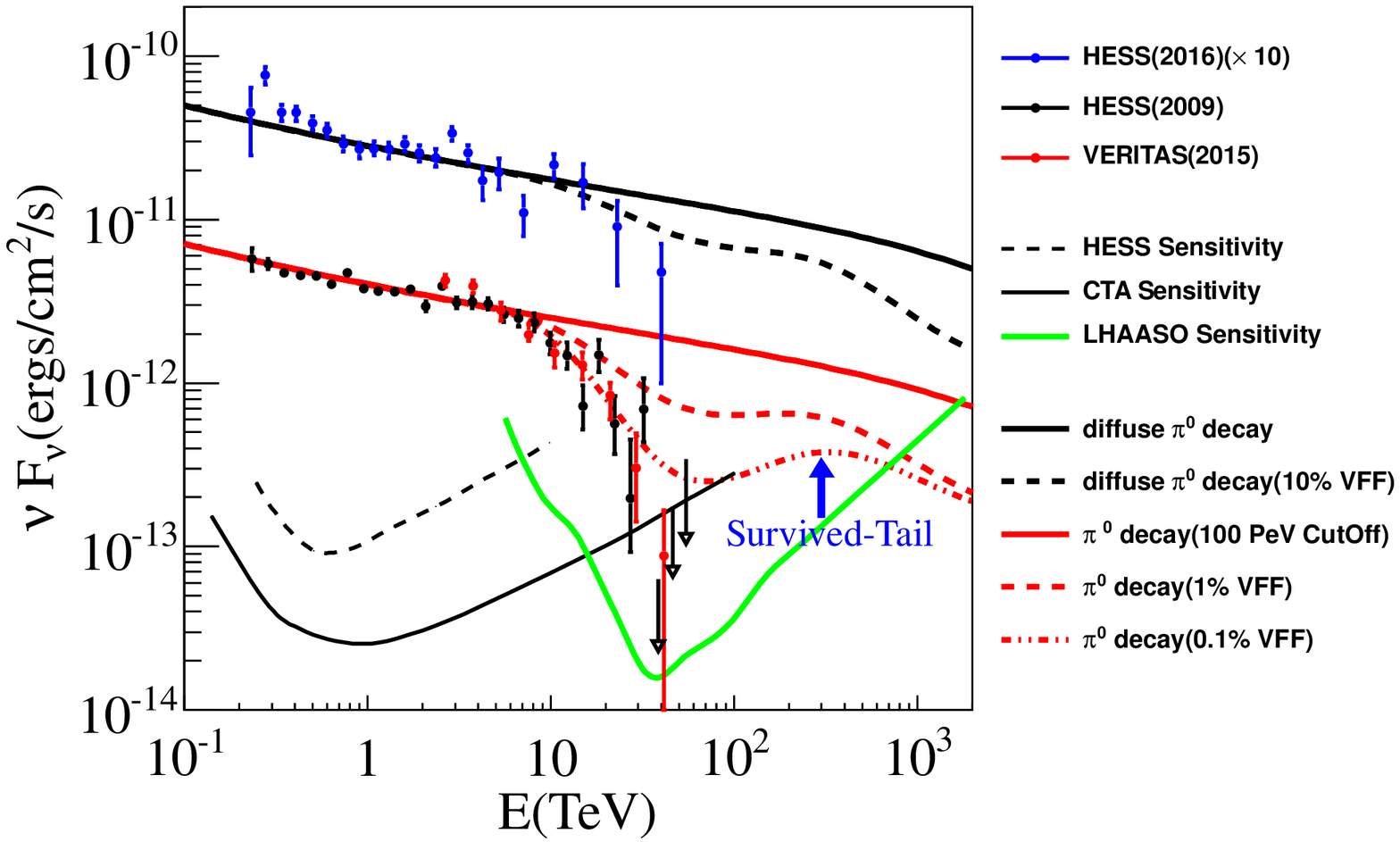}
\caption{The calculated spectrum after the attenuation by considering different VFF.}
    \label{HerschelFilling}
\end{figure*}

  Fig.\ref{HerschelFilling} shows the attenuation with 
  different VFFs. In this calculation, the injection spectrum of protons is assumed as 
  a power law with a break energy at 4 PeV (left panel of Fig.\ref{HerschelFilling}) and 
  100 PeV (right panel of Fig.\ref{HerschelFilling}). The break energy reflects the maximum energy that 
  protons can achieve in the GC activity. The choice of 4 PeV originates from the knee position of all particle
  spectra and 100 PeV comes from the newest observation of light nuclei \citep{2016Natur.531...70B}. 
  The spectral index is assumed to be 2.3 for both point and diffuse one. 
  The total energy $3.08\times 10^{48}$ and $2.16 \times 10^{48}$ erg is fixed for the point and diffuse one respectively. 
  The gas density $n_{gas}=10^3 cm^{-3}/f_V$ is adopted after considering the VFF, which
  keeps the same amount as the above calculation with an average density of $10^3 cm^{-3}$.
  It is clear that the attenuation effect can be significantly enlarged
  in case of the VFF. Taking into account the newly estimated photon density and by 
  adopting a VFF of about $0.1\%$, the observed spectrum from the point source of GC can be well described.
  Away from the GC, the VFF will grow, which leads to the weaker of ISRF. In the diffuse emission of the GC,
  a similar calculation is performed with $1\%$ VFF, which is roughly consistent with the observation. 
  The typical features of a surviving tail is expected at $\sim$100 TeV for the point source.
  We hope that the high precise measurements of the $\gamma$-ray spectrum from TeV to hundreds of TeV 
  will be performed by future projects, such as CTA \citep{2011arXiv1111.2183C} and LHAASO \citep{2010ChPhC..34..249C},
  and can give the ultimate answer to our model.

\subsection{Neutrino emission}

  When the observed $\gamma$-rays are mainly from the decay of the neutral pions which are 
  the products of hadronic interaction between CRs and the ambient gas, a similar 
  amount of neutrinos are expected to be produced from the charged pion decay. 
  The $\gamma$-ray spectrum may be distorted by the absorption interaction, and neutrinos 
  can carry the spectrum of the parent CR interaction. The neutrino spectrum can thus provide 
  decisive information to distinguish whether intrinsic acceleration or absorption of the ISRF 
  should be responsible for the cutoff spectrum of the $\gamma$-rays.

  On average, $PP$-collisions an produce equal number of neutral pion and charged pion. Each neutral 
  pion decays to a pair of $\gamma$-rays and each charged pion decays into two muon neutrinos and 
  one electron neutrino. The initial neutrino flux ratio is
  approximately $\nu_e:\nu_\mu:\nu_\tau$ =
  1:2:0 from charged pion decay. However, the flavor ratio is close to $\nu_e:\nu_\mu:\nu_\tau$ = 1:1:1
  at the Earth after vacuum oscillation through traversal of astrophysical distances. So the typical 
  energy of the neutrino($\nu + \bar {\nu }$) coming from charged pion decay is $\sim$0.5
  of the $\gamma$-ray energy from neutral pion decay. 

  High-energy neutrinos can be detected by neutrino telescopes which use either ice or water as 
  target and detector medium. Neutrinos undergo charge current or neutral current interaction 
  with target matter and produce leptons inside the detector (as a contained event) or in the vicinity of 
  the detector (through-going event). The high-energy muons can generate Cerenkov light while electrons 
  and tau particles may develop to shower which can also generate Cerenkov light for further detection.

  There are two modes of muon event rates: one is the contained event, and the other is
  the through-going event. 
  The contained event is described as the interaction for neutrinos with nucleons inside the
  detector and given by \citep{1996APh.....5...81G,1998PhRvD..58i3009G,2005PhRvD..71i3010G,2006PhRvD..74f3007K}
\begin{equation}
\left ( \frac{dN_\mu}{dE_\mu}\right )_{con} =  k V_{det}\frac{d\Phi _\nu}{dE_\nu} 
e^{E _ \nu/E _ \nu ^ {cut}} \sigma_{CC}(E_\nu)e^{-\tau}
\end{equation}
  where $V_{det}$ is the detector volume, which is adopted to be 1 km$^3$; $E_\nu^{cut}$ is the
  high-energy cutoff of the neutrino spectrum; the term $k$=$N_A\rho T<1-y(E_\nu)>^{-1}$ takes into account
  observation time (T), normalization of the muon spectrum, and the molar density of water (KM3Net) or ice (IceCube).
  The through-going event is described as the interaction for neutrinos with nucleons outside the detector and is given
  by \citep{1996APh.....5...81G,1998PhRvD..58i3009G,2005PhRvD..71i3010G,2006PhRvD..74f3007K}
\begin{equation}
\left ( \frac{dN_\mu}{dE_\mu}\right )_{thr} = \frac{N_A\rho TA_{det}}{\alpha +\beta E_\mu}
\times \int \limits _{E_{\mu}}^{+\infty} dE_\nu \frac{d\Phi _\nu}{dE_\nu} e^{-E_\nu/E^{cut}_\nu}
\sigma _{CC}(E_\nu)e^{-\tau}
\end{equation}

\begin{figure}[!htb]
\centering
\includegraphics[width=0.48\textwidth]{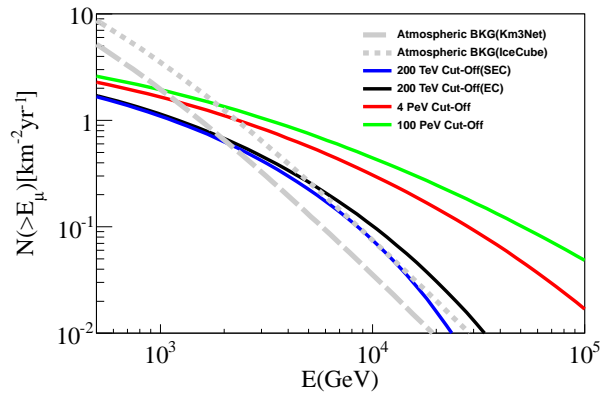}
\caption{The observation of total muon number in one year above an assumed muon energy by KM3Net experiment.}
    \label{neutrino}
\end{figure}

  Based on the above formula, the total muon event number is calculated for the KM3Net experiment.
  As shown in Fig.\ref{neutrino}, it is obvious that the KM3Net has the potential ability to
  observe GC neutrinos with a few years of operation when the break energy of the protons is more than PeV.
  On the contrary, if the break energy of the protons is at $\sim$200 TeV, the GC neutrino events can not
  be separated from atmospheric neutrino background. 
  The observation years to reach a 3$\sigma$ significance level for different cases are estimated and listed in
  Table 1.
\begin{center}
\begin{table}[h]
\begin{center}
\caption{Comparison between the expected signal and the atmospheric neutrino background for different break energy of proton\\}
\begin{tabular}{|c|c|c|c|c|}\hline
  Mode & $E_{th}(TeV)$ & $N_{\mu +\bar {\mu}}$ & $N_{\mu +\bar {\mu}}^{atm}$ & yrs(3$\sigma$)  \\\cline{1-5}
200 TeV energy    &     1         &        1.04           &        1.90      &  29.6         \\\cline{2-5}
       cut-off    &     5         &        0.26           &        0.13      &  35.2         \\\cline{2-5}
                  &     10        &        0.11           &        0.03      &  56.4        \\\hline
4 PeV energy      &     1         &        1.65           &        1.90      &  12.1        \\\cline{2-5}
cut-off           &     5         &        0.57           &        0.13      &  10.0        \\\cline{2-5}
                  &     10        &        0.3            &        0.03      &  6.9         \\\hline
100 PeV energy    &     1         &        1.92           &        1.90      &  8.9         \\\cline{2-5}
cut-off           &     5         &        0.75           &        0.13      &  5.1         \\\cline{2-5}
                  &     10        &        0.45           &        0.03      &  5.7         \\\hline

\end{tabular}
\end{center}
\end{table}
\end{center}

\section{Discussion}
 
  The open question is how to distinguish the production mechanism of the $\gamma$-ray 
  cutoff between the intrinsic acceleration ability of CRs and the attenuation of the ISRF.
  One possible way is to observe the typical feature of the surviving tail.
  The other effective way is to find an instance of the clumpy structure. 
  If the line shape of the calculated spectrum of $\gamma$-ray emission in the clumpy structure 
  is consistent with the result observed 
  by the HESS experiment, this can further support our model of adopting the VFF.

\begin{figure}[!htb]
\centering
\includegraphics[width=0.48\textwidth]{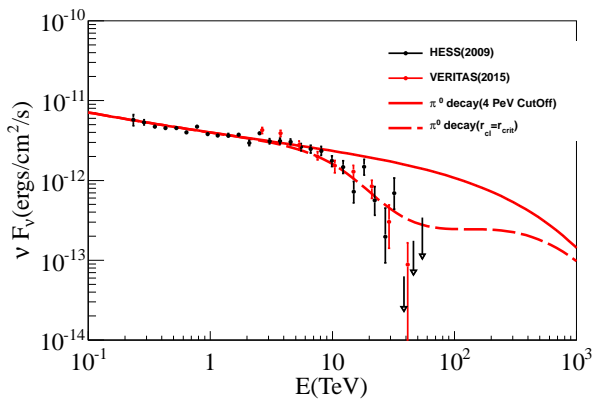}
\caption{The calculated spectrum after the attenuation of Clump 3 with YSO candidate 526817 in it.}
    \label{bumpAlma}
        \end{figure}

  In the central region, the minispiral is a region with a stellar population cluster and
  density structure, which consists of four
  main components: the northern arm, the western arc, the eastern 
  arm and the bar \citep{2012A&A...538A.127K}. In those streamers, it is very bright in
  the near-infrared wave band and possible $PP$-collision regime.
  Recently, ALMA has also observed some separated clumpy structures \citep{2013ApJ...767L..32Y}.
  We take clump 3 as an instance of the clumpy structure to estimate the attenuation of high-energy $\gamma$-rays.
  The clumps in the vicinity of the GC are exposed to strong tidal forces that tend to disrupt the
  clouds, except that the self-gravity is large enough to overcome the tidal shear.
  The tidal limit for the clump mass $M_{cl}$ and the clump radius $r_{cl}$
  is given \citep{1987ApJ...312...66M,2001A&A...367...72V}:

\begin{equation}
\frac{3}{5}\frac{GM_{cl}^2}{r_{cl}} \geq \frac{1}{5}M_{cl}r_{cl}^2|f^\prime({R})|
\end{equation}
  where $f^\prime(R)$ is the derivative of $f(R)$, $f(R)=\frac{GM(R)}{R^2}$ and 
  $R$ is the clump’s distance to the GC.
  The mass distribution can be defined as $M(R)=M_0+M_1R^{1.25}$, where $M_0=4\times 10^6M_{\odot}$
  and $M_1=1.6\times 10^6M_{\odot}$pc$^{-1.25}$. Consider $R$=0.12 pc, $M_{cl}\sim$ 30 M$_\odot$,
  the critical clump radius $r_{crit}\sim 1500$ AU. In this calculation, we take the critical 
      radius $r_{crit}$ as the clump size. In addition, the radiation field of clump 3 is given by the ALMA
      experiment \citep{2013ApJ...767L..32Y}. The attenuation of $\gamma$-rays can be estimated 
      based on Equation \ref{eqEBL} as shown in Figure \ref{bumpAlma}.
  It is obvious that our calculation of the $\gamma$-ray spectrum
  is consistent with the HESS observation line shape. 
  This further gives the possibility that the enhanced density of photons by the VFF leads to the cutoff of the $\gamma$-ray
  spectrum for the GC point source.

   In addition, the overall behavior of the GC is quite silent now, except for some continuous weak 
      activities \citep{1982ApJ...258..135B,1992ApJ...387..189D,
      2009ApJ...698..676D,2011ApJ...728...37D,2016MNRAS.456.1438Y}. It is obvious that
      such weak activities cannot supply enough energy to satisfy the power requirement of
      the $\gamma$-ray emission observed by the HESS experiment. However, there is sufficient evidence to prove that
      the GC had the violent activities in the past, such as X-ray outbursts \citep{2013A&A...558A..32C} and 
      Fermi-Bubbles \citep{2010ApJ...724.1044S}. The HESS Collaboration also proposed that the activity should operate for about
      1000 years to satisfy their observations. In our calculation, we also think that such past activity supplied the required
      the power required to accelerate the protons to PeV energy.

\section{Conclusion}

  The GC is a unique laboratory for studying the origin, acceleration, and propagation of 
  CRs. Considering the inhomogeneous distribution of the ISRF in the 
  GC, $\gamma$-ray absorption is found to be enhanced largely. 
  If the VFF of the clumpy structure is assumed to be 0.1$\%$, 
  the absorption of the $\gamma$-rays can lead to the sharp
  cut-off at about tens of TeV and
  a "survived-tail" at about 100 TeV and sharp cutoff for $\gamma$-ray spectrum are expected.
  Away from GC, the VFF grows up and the attenuation becomes less important.
  The "surviving tail" as the tagged feature can be observed by future projects, such as CTA and LHAASO. 
  High-energy neutrino detection 
  is crucial in distinguishing whether the absorption or the intrinsic acceleration is the cause of 
  the $\gamma$-ray spectrum cutoff. 
  If our model is right, the KM3Net experiment will reach a 3 $\sigma$ observation for multi-TeV muon track neutrinos
  in about $5\sim 10$ years of observation. 
  Owing to the higher background numbers of atmospheric neutrinos for IceCube than KM3Net,
  the sensitivity to GC region for IceCube is a little lower than for KM3Net \citep{2014ApJ...796..109A}.
  More years of operation would be required for the IceCube experiment to reach a 3$\sigma$ significance level of observation.

\section*{Acknowledgements}
  We thank Prof. Hong-Bo Hu and Qiang Yuan for helpful discussion.
  This work is supported by the Ministry of Science and Technology of
  China, Natural Sciences Foundation of China (11405182, 11135010, 11663006, 11647311)

\emph{Note added}:At the same time, a similar study was submitted that draws a consistent conclusion
  concerning gamma ray absorption \citep{2016arXiv160408791C}.
\bibliographystyle{apj}
\bibliography{gc}

\end{document}